
\input phyzzx

\PHYSREV

\titlepage
\title{
Lorentz-Invariant ``Elements of Reality" and the Question of
Joint Measurability of Commuting Observables}
\author{Lev Vaidman}

\address{
School of
Physics and Astronomy \break
Raymond and Beverly Sackler Faculty of Exact Sciences \break
Tel-Aviv University, Tel-Aviv, 69978 ISRAEL\break
BITNET: LEV@VM.TAU.AC.IL}

\abstract

 It is shown that the joint measurements of some physical variables
corresponding to commuting operators performed on pre- and
post-selected quantum systems invariably disturb each other. The
significance of this result for recent proofs of the impossibility of
realistic Lorentz invariant interpretation of quantum theory (without
assumption of locality) is discussed.
\vfill
hep-th/9305162 ,  TAUP 2027-93, to be published in PRL.
\endpage

\def \u { \uparrow }
\def \d { \downarrow }

Recently, few authors \REFS\PI {I. Pitowsky, \journal Phys. Lett.
&A156, (91) 137.} \REFSCON\CPP {R. Clifton, C. Pagonis, and
I. Pitowsky,  {\it Philosophy of Science Association} 1992, Volume 1
pp.114.} \REFSCON\HA {L. Hardy
\journal Phys. Rev. Lett. &68, (92) 2981.} \REFSCON\CN {R. Clifton,
P. Niemann
\journal Phys. Lett.
&A166 (92) 177.} \refsend
 by taking very
plausible definition of
``elements of reality" claimed to show that Lorentz invariant realistic
interpretation of quantum mechanics is not possible.  Contrary to the
Einstein-Podolsky-Rosen (EPR) argument, their proof have not based on
the locality assumption.
In this Letter we will show that contradiction disappear when we
abandon a ``product rule" of elements of reality, even for elements of
reality corresponding to commuting operators.  We will show that the
product rule has to be abandoned because joint measurements of
commuting operators in the considered situations invariably disturb
each other.

The plan of this Letter is as follows: We shall start with the
discussion of the problem of measurement performed on a pre- and
post-selected quantum system.  We will show that even commuting
operators can not be measured on such systems without disturbing each
other.  Then we shall briefly present the arguments against Lorentz
invariant realistic interpretation of quantum mechanics due to
Pitowsky\rlap,\refmark{\PI -\CPP}\ and Hardy\rlap,\refmark{\HA -\CN}\
and we shall explain their usage
of the ``product rule".  We shall conclude with a brief discussion of
the two-state vector approach in which the elements of
reality are Lorentz invariant.

In every textbook of quantum mechanics we can find a condition
for simultaneous measurability of variables $A$ and $B$: the
corresponding operators must commute:
$$[A,B] = 0. \eqno(1)$$
Commutativity of the operators $A$ and $B$ is a strong sufficient
condition, in fact, for a given quantum state $|\Psi \rangle$ it is
enough to require ``commutativity on a state":
$$[A,B] |\psi \rangle = 0. \eqno(2)$$
Commutativity condition (2) is sufficient and necessary condition for
simultaneous measurability of $A$ and $B$.  If, the operators $A$ and
$B$ do not commute, the measurement of one disturbs the outcome of the
other.  There is correspondence between unmeasurability of both
variables and the fact that standard formalism of quantum theory can
not associate well-defined values to two noncommuting operators.

One other way to see this property is to consider the standard
measuring procedure\rlap.\Ref\VN{
J. von Neumann,  {\it Mathematical Foundations of Quantum Theory}
(Princeton, New Jersey:  Princeton University Press, 1983.).
}\ The interaction Hamiltonian is given
by
$$H_{int}= g(t) p A,\eqno(3)$$
where $p$ is a canonical momentum of the measuring device; the
conjugate position $q$ corresponds to the position of a pointer on the
device.  The time-dependent coupling $g(t)$ is constant for a short
time interval corresponding to the measurement.  Therfore, during the
time of the measurement we obtain (in the Heisenberg picture):
$${dB \over dt}=i [H,B] =ig(t) p [A,B].
\eqno(4)$$
 Thus, it is clear why variables corresponding to commuting operators
are measurable simultaneously without mutual disturbance, while
measurement of non-commuting operators disturb each other.

 If $A$ and $B$ commute, and if at a given moment we know that
measurement of $A$ must yield $A=a$ while measurement of $B$ must
yield $B=b$ one can safely claim that also the result of a measurement
of $AB$ is known and equal $ab$.
I repeat this well known fact because surprisingly it is not
true when we consider a pre- and post-selected quantum system.

  Let us
spell out what is the pre- and post-selected quantum system.  We
consider a quantum system at time $t$.  For simplicity we choose zero
free Hamiltonian.  At time $t_1 <t$ the system was prepared in a
quantum state $|\Psi_1\rangle$, and at the time $t_2 >t$ another
measurement was performed and it was found in a
state $|\Psi_2\rangle$.  We ask questions about possible measurements
at time $t$.

Consider a measurement of a variable $A$.  If either $|\Psi_1\rangle$
or $|\Psi_2\rangle$ is an eigenstate of $A$, then clearly the outcome
of the measurement is well defined (it is the corresponding eigenvalue
of $A$) and, measuring of commuting variable $B$ before, after, or
even during the measurement of $A$ will not, in principle, disturb the
measurement of $A$.  However, for a pre- and post-selected quantum
system it might be that the result of the measurement of $A$ is
certain even if neither $|\Psi_1\rangle$ nor $|\Psi_2\rangle$ is an
eigenstate of $A$.  This is the case in which a measurement at any
time in the time period $(t_1, t_2)$ of some variables commuting with
$A$, invariably disturb the $A$-measurement.

The simplest example is the setup proposed by Bohm for analyzing the
EPR argument: two separate spin-1/2 particles prepared, at the time
$t_1$, in a singlet state (which has the same form in every basis)
$$|\Psi_1 \rangle = {1\over {\sqrt 2}} (|\u_1 \d_2\rangle - |\d_1
\u_2 \rangle ). \eqno(5)$$
At the time $t_2$ the measurements of ${\sigma_1}_x$ and
${\sigma_2}_y$ are performed and some certain results are obtained. If
at the time $t$, $ t_1 < t < t_2$, the measurement of ${\sigma_1}_y$
was performed (and this is the only measurement performed between
$t_1$ and $t_2$), then the outcome of the measurement is known with
certainty: ${\sigma_1}_y(t) = -{\sigma_2}_y(t_2)$. If, instead, only
a measurement of
${\sigma_2}_x$
was performed at the time $t$, the result of the measurement is also
certain: ${\sigma_2}_x(t) =- {\sigma_1}_x(t_2)$. The operators
${\sigma_1}_y$ and ${\sigma_2}_x$ are obviously commute, but
nevertheless, measuring
${\sigma_2}_x(t)$ clearly disturb
the outcome of the measurement of ${\sigma_1}_y(t)$: it is not certain
anymore. The EPR argument yields that
${\sigma_1}_x(t)$  is known in this case (it is equal to -
${\sigma_2}_x(t)$), but ${\sigma_1}_y$ is uncertain.

Measuring of the product  ${\sigma_1}_y {\sigma_2}_x$,  is, in
principle, different from the measurement of both ${\sigma_1}_y$ and
${\sigma_2}_x$ together.
In our example the outcome of the measurement of the product {\it
is} certain, but it does not equal to the product of the results
which must come out of the measurements of
${\sigma_1}_y$ and ${\sigma_2}_x$ when every one of them is performed
without the other.
 Note that we can rewrite the
operator
of the product as a modular sum: ${\sigma_1}_y {\sigma_2}_x = (
{\sigma_1}_y + {\sigma_2}_x)_{mod4} - 1$ and for such operators there
is a method for instantaneous measurements which
 uses solely local interactions\rlap.\Ref\AAVNL{
Y.Aharonov, Albert, and L. Vaidman \journal Phys.  Rev.& D34, (86)
1805.}

To show this, and for future applications to other pre- and
post-selected quantum systems let us apply the formalism pioneered by
Aharonov Bergmann and Lebowitz \Ref\ABL{Y.  Aharonov, P.G.  Bergmann,
and J.L.  Lebowitz \journal Phys.  Rev. &B 134 (64) 1410.} for
calculating probabilities of the results of measurement performed
between two other measurements.  If the first measurement prepared the
state $|\Psi_1\rangle$, the final measurement found the state
$|\Psi_2\rangle$, then the probability for a result $A=a_n$ is given
by \Ref\AV {Y.Aharonov and L. Vaidman \journal J. Phys. & A 24 (91)
2315.}
$$
prob(A=a_n ) = {{| \langle \Psi_2| {\bf P}_{A=a_n}
|\Psi_1\rangle |^2} \over {\sum_k
| \langle \Psi_2| {\bf P}_{A=a_k}
|\Psi_1\rangle
 |^2}}~~~~.
\eqno(6)$$
where the sum is over all possible eigenvalues of $A$.
The formula immediately yields probability 1 when $|\Psi_1\rangle$ or
$|\Psi_2\rangle$ is an eigenstate, but it also might yield 1 when
neither of the states is an eigenstate, as we proceed to show on
the example presented above.

In our example the state $|\Psi_1 \rangle$ is given by Eq.(5).
Consider for concreteness the following results of the final
measurements: ${\sigma_1}_x = 1$ and ${\sigma_2}_y = 1$. Then
the state $|\Psi_2 \rangle = |{\u_1}_x {\u_2}_y \rangle $.
For finding the probability of the outcome of the measurement
of ${\sigma_1}_y$ we have to use the projection operators
$ {\bf P}_{[{\sigma_1}_y = 1]} = |{\u_1}_y \rangle \langle {\u_1}_y
|$,
$ {\bf P}_{[{\sigma_1}_y = -1]} = |{\d_1}_y \rangle \langle {\d_1}_y
|$.
Applying all this to formula (6), we, indeed, obtain prob[
${\sigma_1}_y = -1] = 1$. In the same way we obtain
prob[${\sigma_2}_x= -1] = 1$. For calculation of the probabilities of
the measurement of the product ${\sigma_1}_y {\sigma_2}_x$ we shall
use the projection operators
$$\eqalign{
 {\bf P}_{[{\sigma_1}_y {\sigma_2}_x  = 1]} =&
 |{\u_1}_y {\u_2}_x  \rangle \langle {\u_1}_y {\u_2}_x| +
 |{\d_1}_y {\d_2}_x  \rangle \langle {\d_1}_y {\d_2}_x|,\cr
 {\bf P}_{[{\sigma_1}_y {\sigma_2}_x  = 1]} =&
 |{\u_1}_y {\d_2}_x  \rangle \langle {\u_1}_y {\d_2}_x| +
 |{\d_1}_y {\u_2}_x  \rangle \langle {\d_1}_y {\u_2}_x|.
\cr}\eqno(7)$$
The Eq.(6) yields, then,
prob[$ {\sigma_1}_y {\sigma_2}_x= 1] = 0$ contrary to the
consequence of the product rule according to which $
{\sigma_1}_y {\sigma_2}_x= 1$ with probability 1. It follows that the
value of the product
$ {\sigma_1}_y {\sigma_2}_x$
is certain, but it equals to $-1$.

Let us state the main result: for pre- and post-selected quantum
system it might be the case that the operators corresponding to two
observables $A$ and $B$ commute $[A,B] = 0$, and the value of $A$ is
well defined (the outcome of a measurement, if it is the only one to
be performed, is certain), but measuring $B$ invariably disturbs the
results of the measurement of $A$.  Therefore, for pre- and
post-selected quantum system one cannot apply frequently used
``product rule" which is: if it is known with certainty that $A=a$ and
$B=b$, then $AB = ab$.  In fact it might be that the value of $AB$ is
also known with certainty, but it does not equal to $ab$.

Let us turn now to presenting the arguments against possibility of
Lorentz invariant realistic interpretation of quantum
mechanics\rlap.\refmark{\PI -\CN}\ Starting point of these arguments
were the definition of elements of reality and the principle of
Lorentz invariance.  Contrary to the usual EPR-type arguments, no
locality principle, i.e., the impossibility of the action at a
distance, were assumed.  In the discussed works there were adopted the
following definitions: \REF\RED{M.  Redhead, {\it Incompleteness,
Nonlocality, and Realism} (Clarendon, Oxford, 1987).} \item{(i)} {\it
Element of reality}\refmark{RED}:``If quantum predictions dictate with
certainty what the result of measuring a physical quantity at some
time would be then, whether or not the prediction is actually
verified, there exist an element of reality at that time corresponding
to this physical quantity and having equal to the predicted
measurement result."

\item{(ii)} {\it The Principle of Lorentz invariance}:
``If an element of reality corresponding to some Lorentz
invariant physical quantity exists and has a value within space-time
region R with respect to one space-like hyperplane containing R, then
it exists and has the same value in R with respect to any other
hyperplane containing R."

In the usual EPR argument the element of reality corresponding to the
outcome of a measurement fixed just by the {\it possibility} to infer
this outcome from the results of measurements in a causally
disconnected region. Contrary to this, in the present approach, which
does not rely on the locality assumption,  the elements of
reality are fixed by actual measurements performed in a space-like
separated region.

The first argument \refmark{\PI -\CPP} is based on the modified
Greenberger-Horne-Zei-\break linger\Ref\GHZ{
D.M. Greenberger, M. Horne, A. Zeilinger, in {\it Bell's Theorem,
Quantum Theory, and Conception of the Universe}, edited by M. Kafatos}
 (GHZ) setup for proving the nonexistence of local hidden
variables.
Three spin-1/2 particles located
in the corners of a very large triangle move fast in the directions
pointing out of the center of the triangle.  At the time $t_1$ the
particles are prepared in the state:
$$
|\Psi_1 \rangle =
|GHZ \rangle = {1\over {\sqrt{3}}} ({|\u_1}_z {\u_2}_z {\u_3}_z
\rangle - {\d_1}_z {\d_2}_z {\d_3}_z \rangle). \eqno(8)$$
At the time $t_2$ the the spin components in $x$ direction are measured
on all particles and the results
${\sigma_i}_x = x_i$ are obtained. Consider now some possible
measurements performed on the particles at the
 time $t$, $t_1 < t < t_2$. For the three
observers who perform the ${\sigma_i}_x$ measurements (at the rest
frame time $t_2$), the measurements on the other particles (rest frame
time  $t$) are performed {\it after} their
${\sigma_i}_x$ measurement, and they can predict (each in his Lorentz
frame) the following result with certainty:
$$ {\sigma_2}_y {\sigma_3}_y = x_1 \eqno(9a)$$
$$ {\sigma_1}_y {\sigma_3}_y = x_2 \eqno(9b)$$
$$ {\sigma_1}_y {\sigma_2}_y = x_3 \eqno(9c)$$
Eqs.(9a-c) represent elements of reality
in certain space-time regions corresponding to certain Lorentz frames.
The principle of Lorentz invariance yields that these are also the
elements of reality in the rest frame.
 Multiplying  Eqs. (9b) and (9c) we  obtain:
$${\sigma_1}_y^2 {\sigma_3}_y {\sigma_2}_y  = x_2 x_3 \eqno(10)$$
Taking in account that ${\sigma_1}_y^2$ is an identity operator for
spin variables of particle 1, we conclude that
$x_1 = x_2 x_3 $.
The obtained equation, however, contradicts quantum mechanics: in GHZ state
it must be that
$x_1 x_2 x_3 = -1$.

Another example uses just two particles: electron and positron.  Hardy
have used two entangled setups proposed by Elitzur and Vaidman (EV)
for interaction-free measurements\rlap.\Ref\EV{ A. Elitzur and L.
Vaidman, {\it Found.  Phys.} to be published.} They proposed to place
a Mach-Zehnder interferometer tuned to zero counts of one of the
detectors in such a way that the particle's trajectory of one arm of
the interferometer passes through the observed region.  One EV device
tests the point $\cal P$ with a single electron, while the other tests the
same point $\cal P$ at the time with a single positron.  If both electron
and the positron come to the point $\cal P$ together then they annihilate,
and it might happen that both devices yield that the point $\cal P$ is not
empty.  This is the case considered in Hardy's
work\rlap.\refmark{\HA}\ Clifton and Niemann\refmark{\CN} considered a
variation of this idea using Stern-Gerlach devices rather than
interferometers.

Consider the Lorentz frame in which the observer of the electron EV
device has its result first.  She infers
\Ref\F{She {\it retrodicts}, since the events she infers were in her
absolute past.  Hardy succeeds to consider the observer's {\it
predictions} by considering the question: ``is the particle in the arm
of the interferometer which includes $\cal P?$" instead of the
question:`` is the particle in $\cal P$?" See also Ref.(\CN).}
 that the {\it positron} was
in $\cal P$. In other Lorentz frame, however, the observer of the positron
EV device was the first to obtain the result. He deduces that the {\it
electron} was in $\cal P$ at that time.
The principle of Lorentz invariance yields that there are two elements
of reality: the electron in $\cal P$ and the positron in $\cal P$. The product
rule here is very natural: if the electron in $\cal P$ and the positron in
$\cal P$ then the electron and the positron in $\cal P$. The latter,
however, leads to
contradiction since the particles in $\cal P$ have to annihilate and
cannot be detected by the observers.

Both Hardy and Pitowsky obtain their elements of reality as {\it
predictions} of differment observers, but their arguments run only
when they consider ``predictions" of all observers.  However, there is
no Lorentz observer for which all the predictions are inferences from
the past toward the future: at least some of the inferences must be
{\it retrodictions}.  In fact, in both cases we have a quantum system
on which two complete measurement are performed in succession and the
claims about the elements of reality are made for the time in between
these two measurements.  In the example of Pitowsky it is the
preparation of $|GHZ \rangle$ state and then measurement $x$
components of spin for all particles, while in Hardy's case this is
preparation of the electron-positron state and then detection of
electron and positron in certain detectors.  Therefore, the discussion
in the beginning of this Letter is relevant for both examples.

Consider the first example.  The state
$|\Psi_1 \rangle$ is given by Eq.(8); $|\Psi_2 \rangle =
| x_1, x_2, x_3 \rangle$, i.e., the state with certain $x$ components
of spin.
The operators considered between these two states are:
${\sigma_2}_y {\sigma_3}_y,~ {\sigma_1}_y {\sigma_3}_y,$ and
${\sigma_1}_y {\sigma_2}_y $.

The formalism, Eq.(6), yields (as it should be) the probability 1 for
the outcomes given by Eqs.(9a-c).  But it also shows that the
measurements of commuting operators $ {\sigma_1}_y {\sigma_3}_y,$ and
${\sigma_1}_y {\sigma_2}_y $ disturb each other. Eq.(6) yields that
the probability to find both results (9b) and (9c), when measured
together is just $1\over2$. Again the measurement of the product
differs from the measuring both of the operators separately and the
probability to find the outcome of the product measurement
corresponding to the product of (9b) and (9c) is zero since the
outcome have to be given by Eq.(9a).

In Hardy's example the free Hamiltonian is not zero, it describes the
interaction of the electron and the positron with beam splitters and
mirrors of the interferometers as well as the annihilation of the
electron and the positron in $\cal P$.  Therefore, the state $|\Psi_1
\rangle$ in the formula (6) must be the initial state evolved forward
in time until the time $t$, while the state $|\Psi_2 \rangle$ must be
obtained by evolving the final state backward in time until the time
$t$.  (The time $t$, is the moment when the particles might reach the
point $\cal P$.)  Straightforward calculations shows that Eq.(6)
reproduces Hardy's result: in the conditions specified by Hardy's
example if one was testing:``was the electron in $\cal P$?" -- his
result must be ``yes", if another observer was looking for the
positron in $\cal P$, her answer must be ``yes" too, but if both of
them making the measurements each one will have just the probability
$1\over3$ for answer ``yes" (they will never answer ``yes" together).
Again, the operator considered by Hardy is the product of the two
projection operators and its measurement is not equivalent to two
simultaneous inquiries one about the existence of the electron in
$\cal P$ and another about the existence of the positron in $\cal P$.
The measurement of the product is easy to implement: we have to just
look for the photons created by the annihilation of the
electron-positron pair.  The formalism, i.e.  Eq.(6), yields the
probability zero (contrary to 1 obtained from the product rule).

We believe that Redhead's definition of elements of reality is a
plausible one.  It does not lead to contradiction with Lorentz
invariance if we do not adopt the product rule.  But in the light of
what was shown in the beginning of the Letter it is clear that the
product rule is inconsistent with Redhead's elements of reality.  The
elements of reality are obtained under the assumption that there are no
measurements disturbing the inference of the values of the elements of
reality.  Pitowsky and Clifton state it explicitly in their works:
``For our argument, we shall assume that no such intervening
measurements take place."  But as we showed the measurements of the
operators corresponding to the elements of reality the product of
which they consider do intervene with each other.  So, it is
inconsistent with the way of the definition of the elements of reality
to apply the product rule.  If there is an element of reality that
$A=a$ and there is an element of reality $B=b$, it does not follow
that there is an element of reality $AB =ab$.  It might be that the
product $AB$ has a certain value, and therefore it is an element of
reality in the Redhead's language, but it need not to be equal $ab$.

In fact, this is what happens in all the examples considered here.
In the first example we have  elements of reality:``${\sigma_1}_y =
-1$", ``${\sigma_2}_x = -1$", and the product is also an
element of reality, but ${\sigma_1}_y {\sigma_2}_x = -1$. In Pitowsky
example the elements of reality are: ``
$ {\sigma_2}_y {\sigma_3}_y = x_1$\rlap,"
$ {\sigma_1}_y {\sigma_3}_y = x_2$";and the product is also known,
$ {\sigma_2}_y {\sigma_3}_y {\sigma_1}_y {\sigma_3}_y = {\sigma_1}_y
{\sigma_2}_y = x_3 $. Nevertheless $x_1 x_2 \neq x_3$,
($x_1 x_2 = - x_3$). And, in Hardy's  example {\bf P}$_{e^-} = 1$,
{\bf P}$_{e^+} = 1$, but {\bf P}$_{e^-}${\bf P}$_{e^-} = 0$, where
{\bf P}$_{e^-}$, {\bf P}$_{e^+}$ are projection operators on the states
``an electron in $\cal P$" and ``a positron in $\cal P$" respectively.

In no-hidden-variables theorems where only the wave function evolving
from the past towards the future is considered (i.e. there is no
Lorentz frame in which some facts are retrodicted), the product rule
or its generalization to any function of commuting operators is
certainly valid\rlap.\Ref\ME{N.D.  Mermin \journal Phys.  Rev.  Lett.
&65, (90) 3373}\ Its validity is based on joint measurability of
commuting operators.  However, in considerations in which
retrodictions (at least in the eyes of some Lorentz observers) are
involved, measurements of commuting operators do disturb each other,
and the product rule contradicts the spirit of the definition of
elements of reality.

Giving up the product rule allows us to extend the concept of elements
of reality.  Since we anyway consider circumstances in which
retrodictions are involved, we propose to include them fully and give
to it the same status as to predictions.  In the examples presented
here predictions were applied to the future events as well as to the
causally disconnected (space-like separated) events, while
retrodictions appeared only for space-like separated events.  We
propose to allow retrodiction to the past also.  It will fit the
Redhead definition of elements of reality with a minor change of
``quantum predictions" to ``quantum inferences".  Then, in the two
spin-1/2 particles example, the observer which measures ${\sigma_1}_x
(t_2) = 1 $ not only infers that ${\sigma_2}_x(t) = -1$ but also that
${\sigma_1}_x(t) = 1$.
 Or, in the example of Pitowsky there are also
elements of reality at the time $t$, before the actual measurements of
spin $x$ components: ${\sigma_1}_x = x_1, {\sigma_2}_x = x_2$, and
${\sigma_3}_x = x_3$.

According to the definition, the element of reality exists ``whether
or not the prediction [inference] is actually verified..."  The first
corresponds to the counterfactual statement: if someone performed a
measurement at the time $t$ (and no other measurements were performed
between $t_1$ and $t_2$) then the outcome of his measurement is
certain (given by the element of reality).  But even the part
corresponding to ``no actual verification measurement" has physical
meaning.  Recently introduced {\it weak measurements} \Ref\AVW{
Y.Aharonov and L. Vaidman \journal Phys.  Rev. & A 24 (91) 2315} can
test the elements reality almost without disturbing the quantum
system.  We need an ensemble of identically pre- and post-selected
systems.  Each system is practically undisturbed by the measuring
interaction (which is a standard measuring procedures with very weak
coupling), but also each individual measurement yields almost no
information.  However, collecting the results on the ensemble we can
find the results of the weak measurement which is called {\it weak
value}.  It has been proven \refmark{\AV} that in all cases there
exist an element of reality i.e., the outcome value of the
measurement, if measured, is certain, the weak value is equal to this
value.

Taking in account acute difficulties in building Lorentz invariant
realistic description of a time evolution of a quantum system using a
single quantum state\Ref\AA{ Y.Aharonov and Albert \journal Phys.
Rev.& D24, (81) 359} evolving from the past toward the future, we
suggest a description of a quantum system using the two state vectors:
one evolving from the measurement in the past and teh other evolving
backward in time from the measurement in the future (relative to a
given time).  The elements of reality defined by prediction and
retrodiction yield Lorentz invariant description of the history of a
quantum system.

   \refout
   \bye